\documentclass{appolb}
\usepackage{epsfig}

\begin{document}
\title{FLUCTUATION OF THE INITIAL CONDITIONS AND\\ 
       ITS CONSEQUENCES ON SOME OBSERVABLES 
\thanks{Presented at IV Workshop on Particle Correlations and 
        Femtoscopy (WPCF2008), Krakow (Poland), September 
        11-14, 2008.} 
}
\author{Y. Hama, R.P.G. Andrade, F. Grassi, W. Qian 
\address{Instituto de F\'{\i}sica/USP, C.P. 66318,  
         05314-970 S\~ao Paulo - SP, Brazil}
\and
        T. Kodama
\address{Instituto de F\'{\i}sica/UFRJ, C.P. 68528, 
         21945-970 Rio de Janeiro - RJ, Brazil}
}
\maketitle
\begin{abstract}
We show effects of the event-by-event fluctuation of 
the initial conditions (IC) in hydrodynamic description of high-energy nuclear collisions on some observables. Such IC produce not only fluctuations in observables but, due to their bumpy structure, several non-trivial effects appear. 
They enhance production of isotropically distributed  high-$p_T$ particles, making $v_2$ smaller there. Also, they  reduce $v_2$ in the forward and backward regions where the  global matter density is smaller, so where such effects  become more efficacious. They may also produce the so-called {\it ridge effect} in the two large-$p_T$ particle correlation. 
\end{abstract}
\PACS{25.75.-q, 24.10.Nz, 24.60.-k, 25.75.Ld}
  
\section{Introduction}
Hydrodynamics is one of the main tools for studying 
the collective flow in high-energy nuclear collisions. 
In this approach, it is assumed that, after a complex process involving microscopic collisions of nuclear constituents, at a certain early instant a hot and dense matter is formed, which would be in local thermal  equilibrium. After this instant, the system would expand  hydrodynamically, following the well known set of  differential equations. 

The initial conditions (IC) for the 
hydrodynamic expansion are usually parametrized as smooth  distributions of thermodynamic quantities and four-velocity. 
However, since our systems are small, important 
{\it event-by-event fluctuations} are expected in the IC of the real collisions. Moreover, each set of IC should presents strongly {\it inhomogeneous structure}. 
Our purpose here is to discuss some of the effects caused by such fluctuating and bumpy IC on observables.  

\section{Event-by-event fluctuating hydrodynamics} 
 
\begin{figure}[bht] 
\begin{center}
\includegraphics*[width=5.5cm]{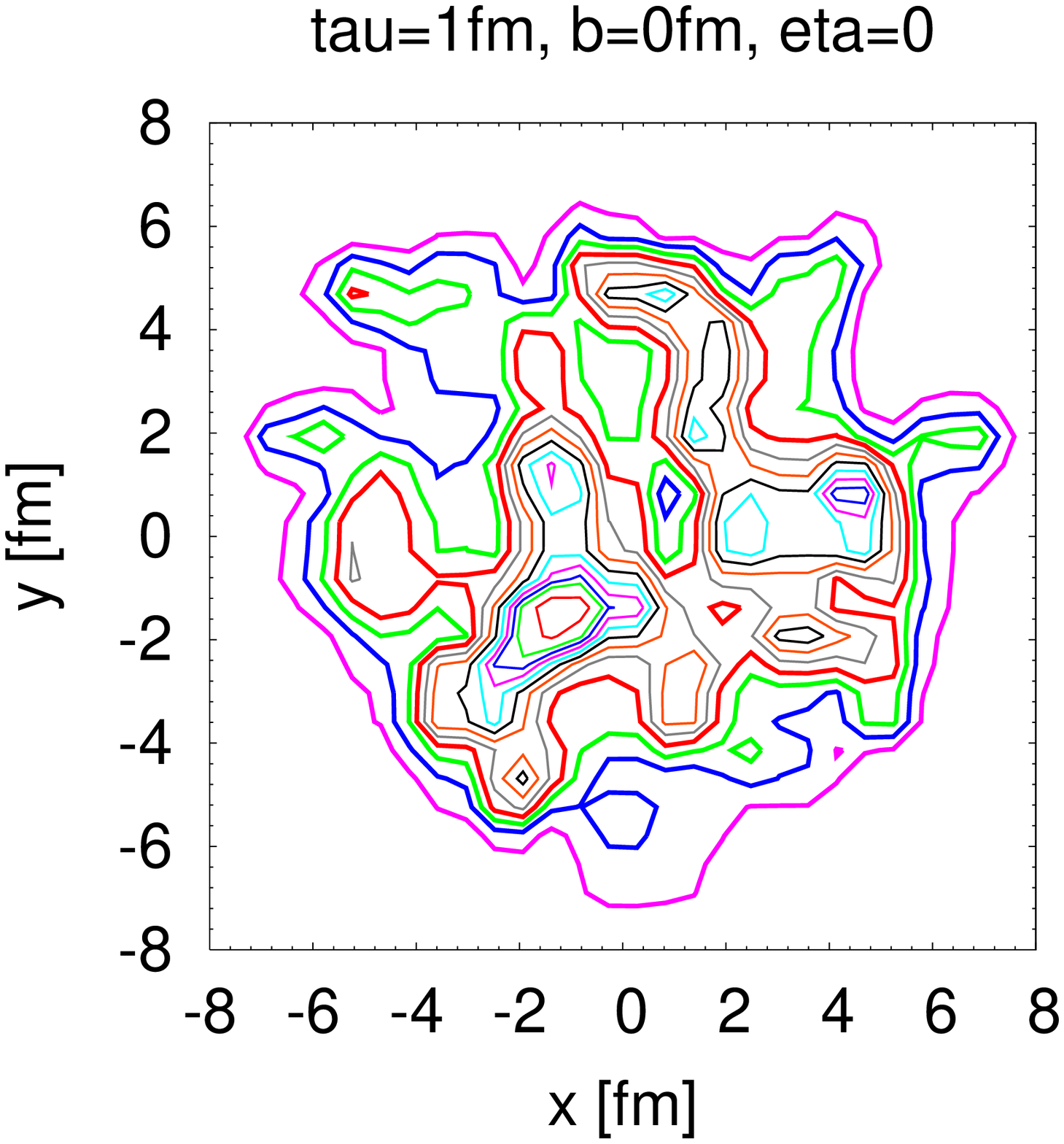} 
\includegraphics*[width=6.5cm]{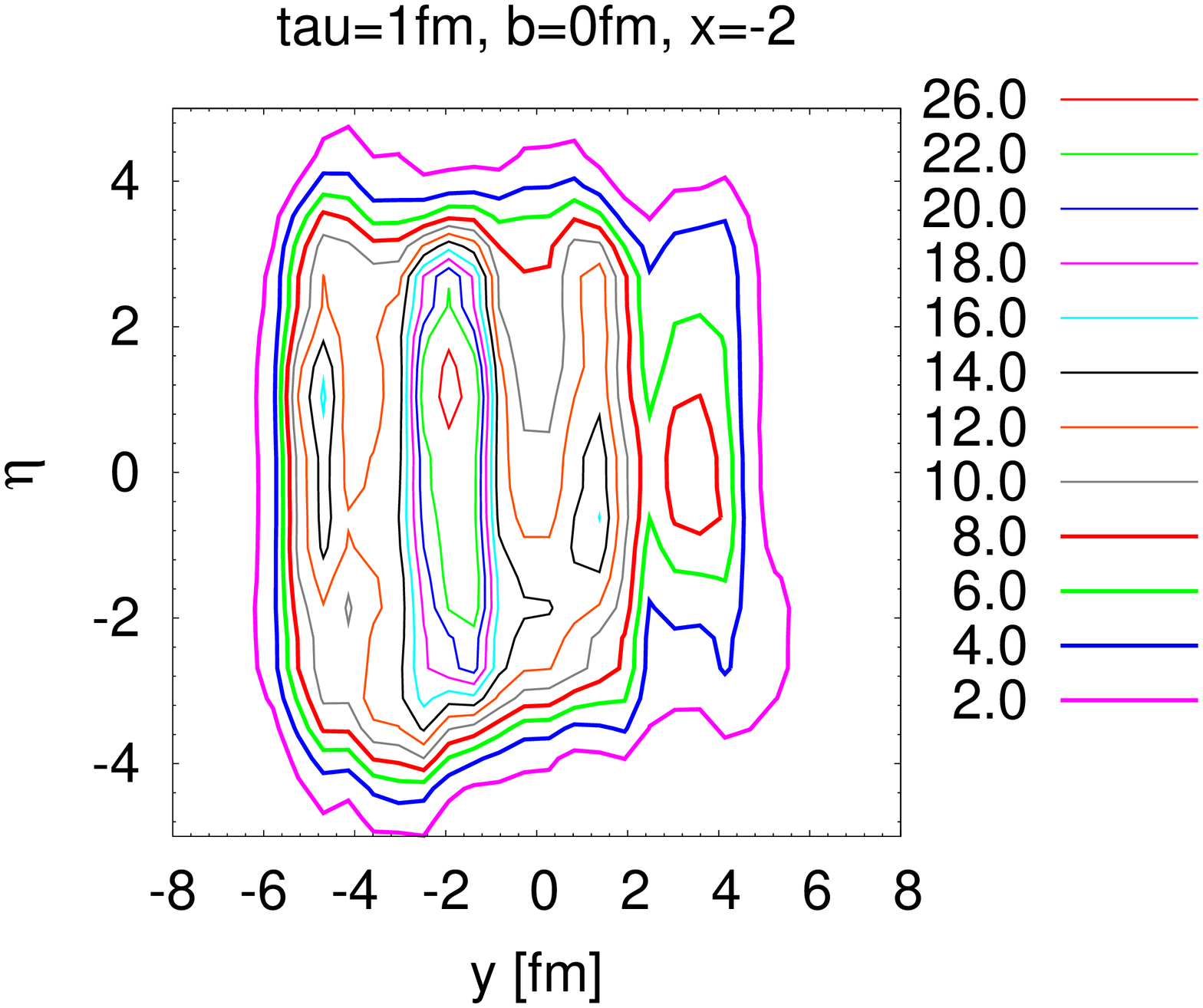} 
\end{center}
\caption{\small Fluctuating IC, produced by NEXUS generator,  
 for the most central Au+Au collisions at 200$\,$A GeV.  
 Left: Energy density distribution is plotted in the $\eta=0$ 
 plane. Right: Corresponding plot in the $x=2$ fm plane.}  
\label{ic} 
\end{figure}

The main tool for our study is called NeXSPheRIO. It is a junction of two  computational codes: NeXus and SPheRIO. 
The NeXus code \cite{nexus} is used to compute the IC. It is a microscopic model based on the Regge-Gribov theory and, once a pair of incident nuclei and their incident energy are chosen, it can produce, in the  event-by-event basis, detailed space distributions of energy-momentum tensor, baryon-number, strangeness and charge densities, at a given initial time  $\tau=\sqrt{t^2-z^2}\sim1\,$fm. 
We show in Fig.\ref{ic} an example of such a fluctuating  event, produced by NeXus event generator, for central Au~+~Au collision at $200\,A\,$GeV. As seen, the energy-density  distribution is highly irregular and in a transverse plane (left panel) it presents several high-density blobs, whereas  in a longitudinal plane (right panel) it presents a baton-like structure. When averaged over many events, these bumps disappear completely giving smooth IC, as those commonly used in hydro calculations. However, this bumpy structure gives  important effects as we will show below.  
The SPheRIO code \cite{sp1,hks}, based on Smoothed Particle 
Hydrodynamics (SPH) algorithm \cite{sph}, is well suited to computing the evolution of such systems, so complex as the one shown in Fig.\ref{ic}. 

\section{Effects of bumpy and fluctuating IC}

In a previous paper \cite{granular}, we discussed some effects of the bumpy structure of IC, as shown in Fig.\ref{ic}, on $p_T$ spectra and $v_2$ coefficient. The main consequence of baton structure\footnote{In that paper, we have considered only the transverse structure of IC and, then called it {\it granular}, but all the discussions remain valid replacing granular by {\it baton} there.} is a violent and cylindrically isotropic expansion of the batons, 
which produces additional isotropic high-$p_T$ particles. 

\subsection{Transverse-momentum spectra}
\begin{figure}[h!tb]
\begin{center}
\includegraphics*[width=7.cm]{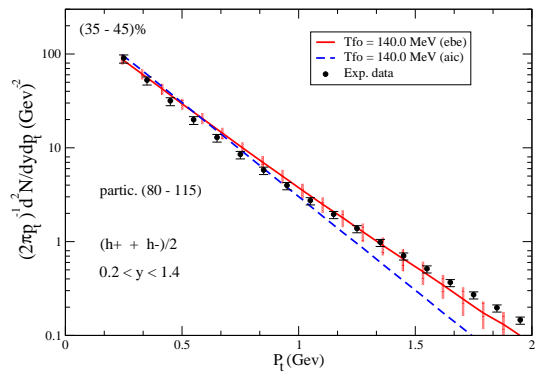}
\end{center} 
\vspace*{-.4cm} 
\caption{Charged-particle $p_T$ distributions (including 
 those from resonance decays) computed in two different ways. 
 The solid line indicates result for fluctuating IC, whereas 
 the dotted line the one for the averaged IC. Data points 
 \cite{PHOBOS1} are also plotted for comparison.} 
\label{dndpt}
\end{figure}

As clearly shown in Fig~\ref{dndpt}, this implies that high-$p_T$ part of the $p_T$ spectrum becomes enhanced as compared with the smooth averaged IC case, making it more concave and closer to data.  

\vspace*{-.5cm} 

\subsection{$p_T$ dependence of $\langle v_2\rangle$}
\begin{figure}[h]
\begin{center}
\vspace*{-.2cm} 
\includegraphics*[width=6.5cm]{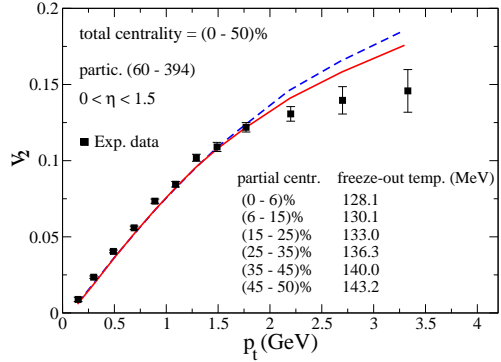}
\end{center} 
\vspace*{-.5cm} 
\caption{$p_T$ dependence of $\langle v_2\rangle$ in the 
 centrality window and $\eta$ interval as indicated, compared 
 with data \cite{PHOBOS2}. The 
 solid line indicates result for fluctuating IC, whereas 
 the dotted one that for the averaged IC. The curves are  
 averages over PHOBOS centrality sub-intervals with  
 freeze-out temperatures as indicated. 
 } 
\label{v2pt}
\vspace*{-.2cm} 
\end{figure}

As for the anisotropy of the transverse flow, we illustrate in Fig.\ref{v2pt} that the elliptic-flow coefficient $v_2$ suffers reduction as we go to high-$p_T$ region, due to the additional high-$p_T$ isotropic component included now. The 
result is closer to the available data. 

\subsection{$\eta$ dependence of $\langle v_2\rangle$}
As for the $\eta$ dependence of $v_2\,$, we know that the average matter density decreases as $\vert\eta\vert$ increases as reflected in the $\eta$ distribution of charged particles, so the isotropic expansion effect of the batons 
becomes more efficacious there and, therefore, $v_2\,$ suffers a considerable reduction in those regions. 
\begin{figure}[h]
\begin{center}
\includegraphics*[width=6.5cm]{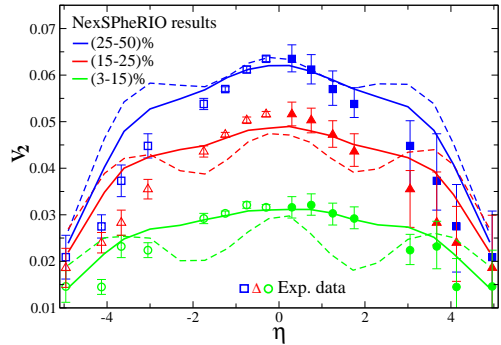}
\end{center} 
\vspace*{-.5cm} 
\caption{$\eta$ dependence of $\langle v_2\rangle$ for three 
 centrality windows. The solid lines indicate results 
 for fluctuating IC, whereas the dotted lines 
 the ones for the averaged IC. Data points\cite{PHOBOS2} 
 are also plotted for comparison. $T_{fo}$ has been 
 taken as in Fig.\ref{v2pt}.} 
\label{v2eta}
\vspace*{-.4cm}
\end{figure} 

\subsection{$v_2$ fluctuation}
\begin{figure}[t] 
\begin{center}
\includegraphics[width=6.5cm]{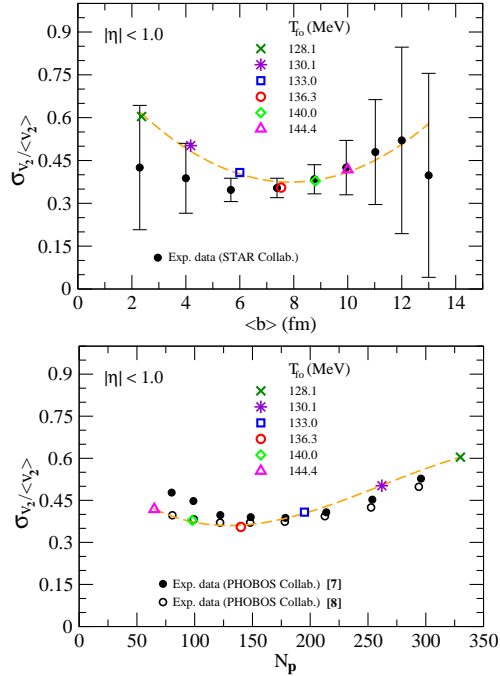}
\end{center} 
\vspace*{-.5cm} 
\caption{$\sigma_{v_2}/\langle v_2\rangle$ computed for 
Au+Au collisions at $200\,A\,$GeV, compared with data. 
Upper: $\sigma_{v_2}/\langle v_2\rangle$ is given as function of the impact parameter $\langle b\rangle$ and compared with the STAR data \cite{STAR}. Lower: the same results are  expressed as function of participant nucleon number $N_p$ and compared with the PHOBOS data, \cite{PHOBOS3}.  
} 
\vspace*{-.5cm} 
\label{flv2}
\end{figure}

The IC fluctuation also implies a large $v_2$ fluctuation. In preliminary works \cite{preliminary} on Au+Au collisions at $130A$ GeV, we showed that this actually happens. The results, with QGP included, have indeed been confirmed in recent experiments\cite{STAR,PHOBOS3}.  
More recent computations for Au+Au at $200A\,$GeV  \cite{fv2} gave similar results. In Fig.~\ref{flv2}, we compare the latter with those data. 
\subsection{Ridge effect} 
\begin{figure}[h!tb]
\begin{center}
\includegraphics*[width=6.cm]{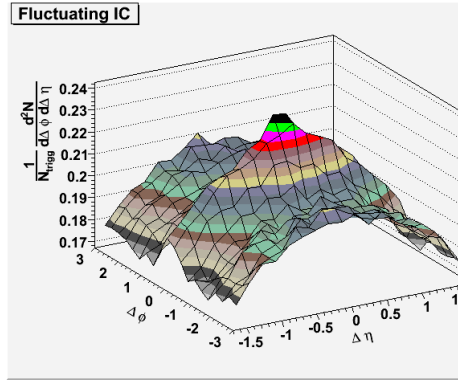}
\end{center} 
\vspace*{-.4cm} 
\caption{Two high-$p_T$ particle correlation, computed with  
 event-by-event fluctuating IC. \cite{jun}} 
\label{ridge}
\end{figure}

Another effect, which is produced naturally by the longitudinal baton structure of IC, as shown in Fig.\ref{ic}, is the so-called {\it ridge phenomenon} which has been experimentally seen in high-$p_T$ nearside correlations \cite{ridge}. Since batons in fluctuating IC which are close to the surface produce longitudinally correlated high-$p_T$  particles, always in the same side of the hot matter, the ridge structure naturally appears. This is shown in Fig.\ref{ridge}, as computed with NeXSPheRIO by J. Takahashi {\it et al.} \cite{jun} The ridge phenomenon has been discussed also in connection with glasma flux tubes \cite{mclerran}.  

\section{Summary} 
In this paper, we presented several consequences of 
event-by-event fluctuating IC with baton structure, computed with NeXSPheRIO code. The main results are: 1) The baton structure of IC produces more concave $p_T$ spectra, as compared with smooth IC; 2) It reduces $\langle v_2\rangle$ both in the high-$p_T$ and large-$|\eta|$ regions; 3) It also produces the {\it ridge structure} in the nearside correlations; 4) Large $v_2$ fluctuations occurs, in good agreement with data.
\vspace*{-.5cm} 
\section*{Acknowledgments}
This work was partially supported by FAPESP, CNPq, FAPERJ and PRONEX. We are grateful to Jun Takahashi for providing us 
with Fig.\ref{ridge} 

\end{document}